# Spectrum-Energy Correlations and *Swift* GRBs

Lorenzo Amati

*INAF – IASF Bologna, via P. Gobetti 101, 40129 Bologna, Italy*

**Abstract.** The correlation between the photon energy at which the redshift corrected $\nu F\nu$ spectrum peaks (hence called "peak energy", $E_{p,i}$) and the isotropic equivalent radiated energy ($E_{iso}$), is one of the most intriguing and debated observational evidences in Gamma- Ray Bursts (GRB) astrophysics. In order to fully exploit this correlation to understand GRB physics and use them as a cosmological tool, we need to understand and reduce selection effects and biases in the sample of GRBs with known redshift. The *Swift* satellite, thanks to its scientific payload and unprecedented slewing capabilities, is providing us new observational evidences and a reduction of selection effects in the process of GRB redshift estimate. I discuss the location of *Swift* GRBs in the $E_{p,i}$-$E_{iso}$ plane and the main implications, with particular emphasis on selection effects, the existence and nature of sub-energetic long GRBs, the GRB-SN connection, the difference between long and short GRBs and the recently discovered sub-class of long GRBs without association with a hypernova. I also briefly review the impact of *Swift* observations on 3-parameters spectrum-energy correlations, which have been proposed as a potentially powerful tool to standardize GRBs for the estimate of cosmological parameters.

## INTRODUCTION

Thanks to the discovery of afterglow emission and the identification of optical counterparts and host galaxies, in the last 10 years it has been possible to estimate the redshift (up t 6.3), and thus the luminosity (up to $10^{53}$ erg/s assuming isotropic emission), of about 100 Gamma-Ray Bursts (GRB). This made possible for the first time to estimate, intrinsic properties of GRBs and search for correlations between them [1]. Among these, the correlation between the photon energy at which the cosmological rest-frame $\nu F\nu$ spectrum peaks ("peak energy", $E_{p,i}$) and the total isotropic-equivalent radiated energy $E_{iso}$ [1] is one of the most robust and intriguing observational evidences in the GRB field. The main implications of the correlation include prompt emission mechanisms, jet geometry and properties, GRB/XRF unification models, identification and nature of sub-classes of GRBs (e.g., sub-energetic, short). See [2] for a review. Also, with the inclusion of an additional parameter (e.g., optical afterglow break time or the "high-signal" time scale) it is a promising tool to standardize GRBs and thus estimate cosmological parameters, see, e.g., [3]. *Swift*, thanks to its high sensitivity and very fast slewing capabilities (and thus the capability of providing localizations with a few arcsec accuracy just a few minutes after the GRB onset), allows a substantial reduction of selection effects in the sample of GRBs with known redshift. This is demonstrated by the different redshift distributions of pre-*Swift* and *Swift* GRBs. Thus, studying the location of *Swift* GRBs in the $E_{p,i}$-$E_{iso}$ plane can give clues on possible selection effects affecting the correlation. Moreover, the unprecedented capabilities of *Swift*, joined with optical follow-up observations and prompt emission measurements by other space experiments (mainly *Konus*/WIND) provided us redshift and peak energy estimates for a few short GRBs and other peculiar events (like the sub-energetic GRB 060218 and the long event with no, or very weak, associated SN, GRB 060614), allowing us to better investigate their nature through their behavior with respect to the $E_{p,i}$-$E_{iso}$ correlation. Finally, the unprecedented sampling of the X-ray afterglow light curves provided by *Swift* is seriously challenging GRB jet models and the $E_{p,i}$-$E_\gamma$ correlation, the most popular among 3-parameters spectrum-energy correlations.

## LONG GRB

Although the $E_{p,i}$-$E_{iso}$ correlation is the tightest and more significant among the empirical correlations between two GRB observables, in the recent years there was a debate, mainly based on BATSE events with unknown redshift, on weather it may be due to

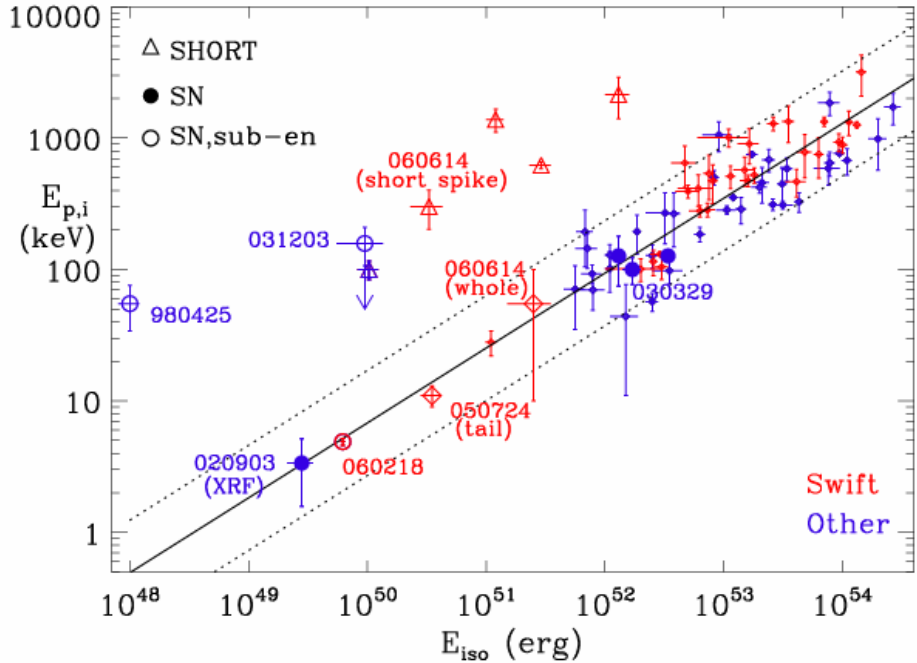

**FIGURE 1.** Distribution of GRBs with known redshift and measured $E_{p,obs}$, as of April 2008, in the $E_{p,i}$-$E_{iso}$ plane. Red symbols correspond to *Swift* GRBs, black ones to events detected by other satellites. $E_{p,i}$ and $E_{iso}$ values are taken from [2], [6] and [14]; the continuous and dashed lines indicate the best fit power-law, and the corresponding $+/-2\sigma$ region, reported by [2]. Short GRBs, GRBs with spectroscopic association with SN, sub-energetic GRBs, XRFs, and other peculiar GRBs are indicated.

substantial selection effects. Different authors came to different conclusions (e.g., [4,5]) but all agreed that Swift measurements would provide a definitive answer, because of the reduced selection effects in its sample of GRBs with known redshift (as detailed in previous section). The drawback is that, due to its limited energy band (15-150 keV), Swift/BAT can provide an estimate of $E_{p,obs}$ only for a small fraction of events (~15%); this reduces the number of GRBs that can be placed in the Ep,i-Eiso plane, the majority of which being those simultaneously detected by *Konus*/WIND. In Fig 1 I show the location of *Swift* GRBs with known redshift and $E_{p,obs}$ in the $E_{p,i}$-$E_{iso}$ plane, as of December 2006. Data are taken from [2,6]. As can be seen, all long GRBs are fully consistent with the $E_{p,i}$-$E_{iso}$ correlation and its scatter as measured before *Swift*. Moreover, as discussed by [6], also those *Swift* GRBs for which only an upper / lower limit to $E_{p,obs}$ can be derived from BAT spectra are potentially consistent with the correlation. These evidences support the hypothesis that the correlation is not affected by significant selection effects. Recently, [7] used a Bayesian approach to estimate $E_{p,obs}$ from BAT spectra for those GRBs having peak energies much above the instrument threshold (~150 keV), finding that a good fraction of events are inconsistent

(specifically above) the pre-*Swift* $E_{p,i}$-$E_{iso}$ correlation. But, in addition to the fact that they assume the BATSE $E_{p,obs}$ distribution, that has been shown by BeppoSAX and HETE-2 measurements to be biased against low $E_{p,obs}$ values, their method is clearly speculative, given that they want to extract information where there are no data and from spectra which are satisfactorily fit by a simple power-law. Clearly, these BAT spectra can be used only to derive lower/upper limits to $E_{p,obs}$ by assuming, e.g., a Band function with beta constrained to the typical value of -2.3. In addition, their Figure 3, left panel, clearly shows that, above ~200 keV, the $E_{p,obs}$ values estimated from BAT by [7] with their Bayesian method are uncorrelated with those estimated with *Konus*/WIND and are consistent with a constant. Thus, the test of the $E_{p,i}$-$E_{iso}$ correlation that they performed is unreliable.

## SHORT GRB

Another great advance occurred in the *Swift* era was given by the first redshift estimates of short GRBs, whose redshift and radiated energy distributions are skewed to lower values with respect to long GRBs. For two of these events, namely GRB 050709 and GRB 051221, it was possible to estimate also $E_{p,obs}$

(from HETE-2 and Konus/WIND, respectively), finding, as can be seen in Fig. 1, that they are inconsistent with the $E_{p,i}$-$E_{iso}$ correlation holding for long GRBs. [6] has also shown that the $E_{p,i}$ lower limits for other few short GRBs with known redshift are inconsistent with the correlation. These evidences further support the hypothesis that the bulk of the prompt emission of short and long GRBs is due to different emission mechanisms. *Swift* also showed that some short GRBs are characterized by a long soft tail, close to the detection threshold but with a total fluence that may be comparable to that of the short spike. Intriguingly, the soft tail of short GRB 050724 [8] is consistent with the $E_{p,i}$-$E_{iso}$ correlation. This indicates that the emission mechanism that produces most of the radiation in long GRBs may be present also in short GRBs, but with much lower efficiency.

## PECULIAR GRB

By joining the capabilities of *Swift* with those of ground optical telescopes and other GRB space experiments, it was possible to detect and study in detail some GRBs with peculiar properties, giving us important clues on GRB origin and physics. Among these, the most interesting cases are those of GRB 060218 and GRB 060614 [9].

### GRB 060218 and Sub-Energetic GRB

GRB 060218 was noticeable because it was very close (z=0.033), very soft, sub-energetic and with a prominent association with a SN (2006aj). These properties make GRB 060218 similar to GRB 980425, which was not only the proto-type of the GRB/SN connection, but also a very peculiar event: much closer than all other GRBs (z=0.0085), sub-energetic, and the only long GRB outlier to the $E_{p,i}$-$E_{iso}$ correlation (Fig. 1). The most popular interpretation of the peculiar properties of GRB 980425 is that this event was a "normal" GRB seen off-axis and that its sub-energetic nature and deviation from the $E_{p,i}$-$E_{iso}$ correlation were due to viewing angle effects. Such an explanation was proposed also for GRB 031203, another sub-energetic GRB with prominent association with a SN and possibly (the $E_{p,i}$ value is uncertain, see Fig. 1) inconsistent with the $E_{p,i}$-$E_{iso}$ correlation too. However, GRB 060218, contrary to the expectations of the off-axis scenarios, is fully consistent with the $E_{p,i}$-$E_{iso}$ correlation. This evidence implies that this event may really be sub-energetic and points to the existence of a population of close and faint GRBs and to a GRB rate which could be much higher than expected before [9].

### GRB 060614 and GRB-SN Connection

GRB 060614 is famous because, despite it is a long event, the limits derived from optical measurements imply that it was originated by a very weak SN or by a different progenitor. Moreover, its light curve is characterized by an initial "short" spike, with properties inconsistent with both the time-lag - luminosity and $E_{p,i}$-$E_{iso}$ relations [10], followed by a longer and soft emission. Thus, it could be the proto-type of a new sub-class of long GRB or a "bridging" event between short and long GRBs. However, as shown by [9], when considering the whole event GRB 060614 is consistent with the $E_{p,i}$-$E_{iso}$ correlation. This implies that the prompt emission properties of long GRBs maybe unrelated to those of the associated SN, given that other (not sub-energetic) GRBs with clear association with a SN (e.g., GRB 060218, GRB 030329, GRB 021211) are consistent too with the correlation. In any case, the fact that GRB 060614 is composed by a short pulse inconsistent with the $E_{p,i}$-$E_{iso}$ correlation and a longer part consistent with it, is similar to what happens for the short GRB 050724 (as discussed above) and may really indicate that there is a "smooth transition" and overlapping between the short and long GRB classes.

## 3-PARAMETERS SPECTRUM-ENERGY CORRELATIONS

The relevance of the $E_{p,i}$-$E_{iso}$ correlation for the physics of GRB and their use as a cosmological tool (e.g., to study the star formation rate evolution, to constrain cosmological parameters) stimulated further search for correlations of $E_{p,i}$ with other GRB luminosity indicators and for three parameters correlations involving the spectral peak energy, a GRB luminosity indicator and other observables. These studies found that $E_{p,i}$ correlates, in addition to $E_{iso}$, with the average isotropic-equivalent luminosity $L_{iso}$ and the isotropic-equivalent peak luminosity ($L_{p,iso}$). This is not surprising, given that these two quantities are strongly correlated with $E_{iso}$. More intriguingly, it was then found that, by including a third observable, the $E_{p,i}$-$E_{iso}$ correlation become more tight, i.e., its intrinsic dispersion (discussed and quantified, e.g., in [6]) reduces of about 50%. The GRB spectrum-energy correlations, the link between them and the observables in play, are schematically summarized in Fig. 2. The first of 3-parameters spectrum-energy correlations was found between Ep,i , Eiso and the break time of the optical afterglow light curve, $t_b$. The $E_p,i$-$E_{iso}$-$t_b$ correlation holds both as it is, in its fully empirical form, and by assuming that the break in the optical afterglow light curve is a signature of collimated emission. In this interpretation, $t_b$ gives (under some assumptions on radiation efficiency and circum-burst environment density profile) the jet opening angle, and thus $E_{iso}$ can be converted into the collimation-corrected (or "true") radiated energy, $E_\gamma$ . Given its low dispersion, the $E_{p,i}$-$E_\gamma$ correlation was proposed as a tool to standardize GRB, in a way similar to SN Ia (see [3] for a review of 3-parameters spectrum-energy correlations and their potential cosmological use). However, there are several issues that, for now, make the community cautious in the use of this correlation for cosmology: the fact that it is based on a much lower number of events with respect to the $E_{p,i}$-$E_{iso}$ correlation (because it requires also the detection of a break in the optical afterglow), the still lacking comprehension of the physics behind it, the "circularity" problem (i.e., given that nearly all GRB lie at z > 0.1, it cannot be calibrated at low distances, where it would not depend on the assumed cosmology, as instead is possible for SN Ia; this could be solved by detecting an high enough number of GRBs with redshifts consistent within ~0.1). Moreover, *Swift* observation put in some trouble the $E_{p,i}$-$E_\gamma$ correlation, because of the discovery of non achromatic afterglow light curve break (contrary to simple jet models predictions), of late X-ray afterglow light curve with no breaks and of possible outliers to the correlations (see [11,12] for discussions). Another promising, because of its low dispersion, 3-parameters spectrum-energy correlation is the $E_{p,i}$-$L_{p,iso}$-$T_{0.45}$ correlation [3] ($T_{0.45}$ is a characteristic GRB time scale linked to the time during which the emission is above a given level with respect to the background). With respect to the $E_{p,i}$-$E_\gamma$ , this correlation has the advantage of being based only on prompt emission properties. However, recent analysis based on updated samples show that its lower dispersion with respect to the $E_{p,i}$-$E_{iso}$ correlation is not confirmed [13].

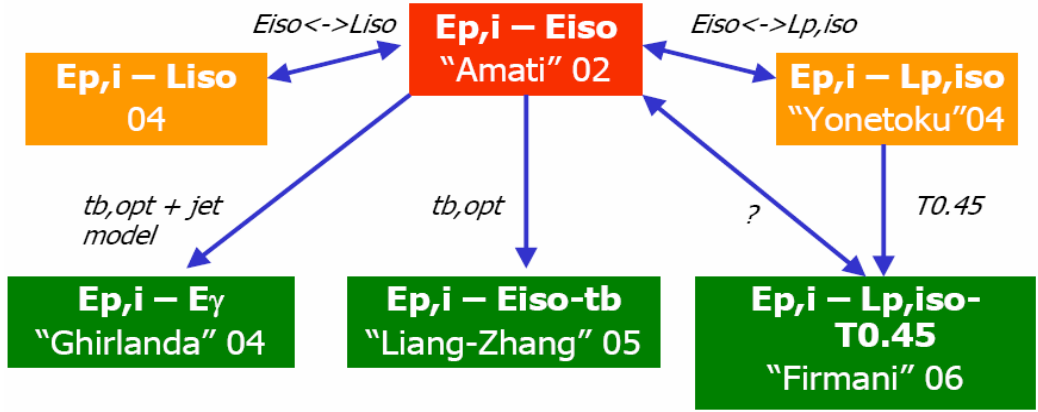

**FIGURE 2.** The "genealogy" of spectrum-energy correlations. The "name" often found in the literature is reported below each correlation, together with the year. Also shown are the link between the correlations (arrows) and the relevant observables.